\pdfoutput=1
\RequirePackage{fix-cm}
\documentclass{svjour3}                     
\smartqed  
\usepackage{graphicx}
\usepackage{bm}
\usepackage{amsmath}
\usepackage{amssymb}
\usepackage{mathptmx}      
\usepackage{latexsym}
\usepackage{verbatim}
\usepackage{color}
\usepackage[english]{babel}
\newcommand{\mmat}[1]{\ensuremath{\boldsymbol{#1}}}

\journalname{Journal of Computational Electronics}
\begin{document}
\title{Negative differential resistance in graphene-nanoribbon--carbon-nanotube crossbars: A first-principles multiterminal quantum transport study}


\titlerunning{Graphene-nanoribbon--carbon-nanotube crossbars}        

\author{Kamal K. Saha \and
        Branislav K. Nikoli\' c
}


\institute{Kamal K. Saha \and
           Branislav K. Nikoli\' c  \at
           Department of Physics and Astronomy, University of Delaware, Newark, DE 19716, USA  \\
              \email{bnikolic@udel.edu}           
}

\date{Received: date / Accepted: date}

\maketitle

\begin{abstract}
We simulate quantum transport between a graphene nanoribbon (GNR) and a single-walled carbon nanotube (CNT) where electrons traverse vacuum gap between them. The GNR covers CNT over a nanoscale region while their relative rotation is 90$^\circ$, thereby forming a {\em four-terminal crossbar} where the bias voltage is applied between CNT and GNR terminals. The CNT and GNR are chosen as either semiconducting (s) or metallic (m) based on whether their two-terminal conductance exhibits a gap as a function of the Fermi energy or not, respectively. We find nonlinear current-voltage ({\em I--V}) characteristics in all three investigated devices---mGNR-sCNT, sGNR-sCNT  and mGNR-mCNT crossbars---which are asymmetric with respect to changing the bias voltage from positive to negative.  Furthermore,  the  {\em I--V} characteristics of mGNR-sCNT crossbar exhibits {\em negative differential resistance} (NDR) with {\em low} onset voltage $V_\mathrm{NDR} \simeq 0.25$ V and peak-to-valley current ratio $\simeq 2.0$. The overlap region of the crossbars contains only $\simeq 460$ carbon and hydrogen atoms which paves the way for nanoelectronic devices ultrascaled well below the smallest horizontal length scale envisioned by the international technology roadmap for semiconductors. Our analysis is based on the nonequilibrium Green function formalism combined with density functional theory (NEGF-DFT), where we also provide an overview of recent extensions of NEGF-DFT framework (originally developed for two-terminal devices) to {\em multiterminal} devices.

\end{abstract}

\section{Introduction}\label{sec:intro}

The discovery of a wonder material {\em graphene}---as the first one-atom-thick crystal whose honeycomb lattice of carbon atoms imposes gapless, massless and chiral Dirac spectrum on its low-energy quasiparticles---has swiftly ignited intense experimental and theoretical studies of its unusual electronic transport properties~\cite{Geim2009}. The ``first wave'' of such studies~\cite{DasSarma2011} has been largely focused on explaining conductivity measurements at small bias voltage applied to large-area graphene samples as a function of carrier density, temperature and magnetic field. These studies have typically relied on simplistic model Hamiltonians.

\begin{figure}
\begin{center}
\includegraphics[scale=0.5,angle=0]{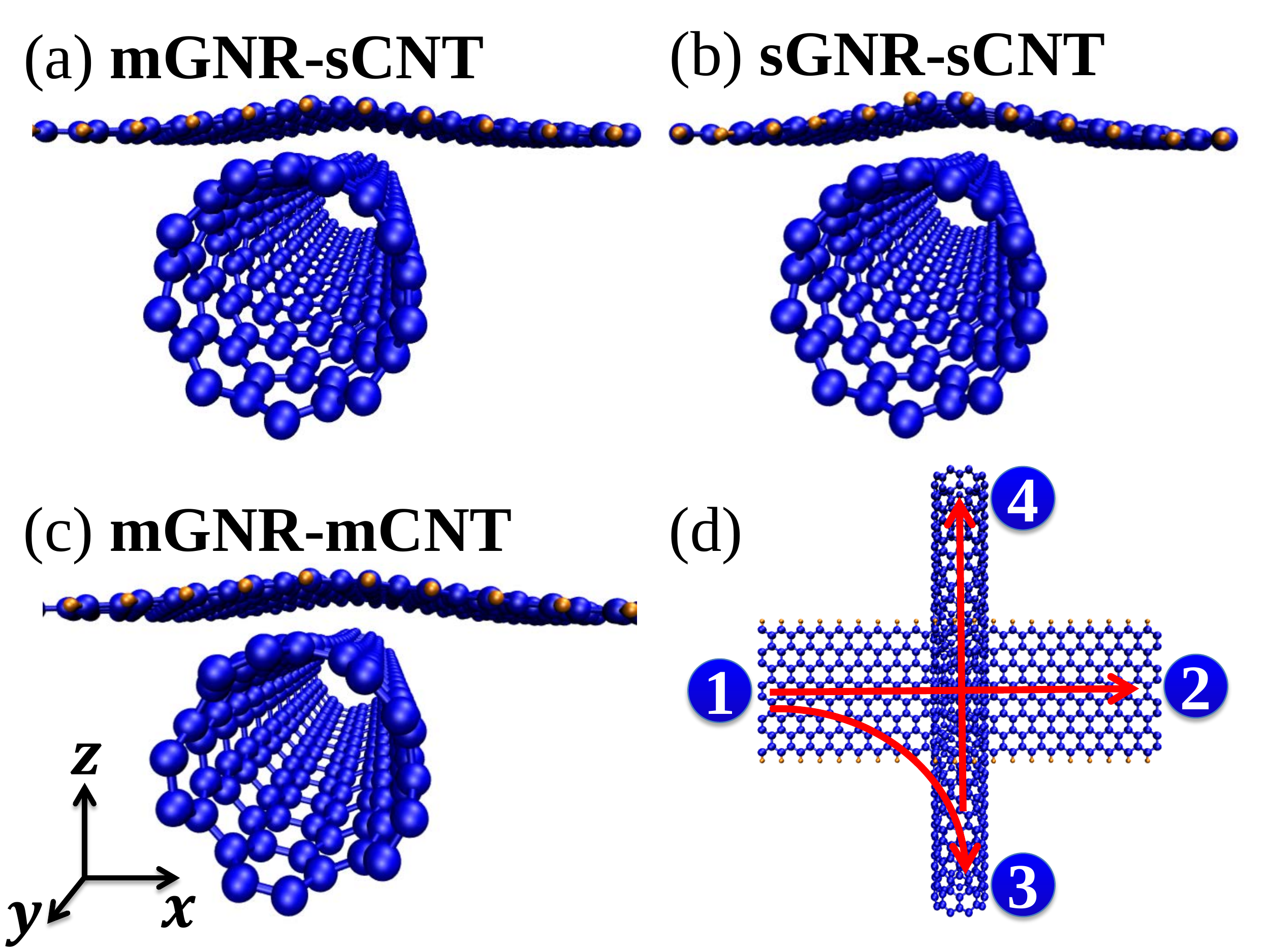}
\end{center}
\caption{Schematic view of GNR-CNT crossbars composed of: (a) metallic 8-ZGNR and semiconducting (9,0)-CNT; (b) semiconducting 14-AGNR and semiconducting (9,0)-CNT; and (c) metallic 8-ZGNR and metallic (5,5)-CNT.  The hydrogen atoms (light yellow) are included to passivate the edge carbon atoms (dark blue) of GNRs. The DFT optimized structures, where Hellmann-Feynman forces acting on ions are less than \mbox{$0.001 \ \mathrm{eV/\AA}$}, show that GNRs in all three panels acquire curvature within the overlap region of the crossbar where the spacing between GNRs and CNTs along the $z$-axis (intersecting with tube axis) is \mbox{$d \approx 3.2$ \AA}. The quantum transport is simulated in the central region consisting of the GNR-CNT overlap region + portion of semi-infinite GNR or CNT electrodes which are labeled in panel (d). Panel (d) also shows some of the possible currents that flow in such {\em multiterminal} geometry, where the focus of our study is on currents like $1 \rightarrow 3$ flowing between GNR and CNT.}
\label{fig:fig1}
\end{figure}

What could be termed the ``second wave'' of studies of electronic transport in graphene has been largely focused on confined structures (such as nanoribbons and nanoflakes~\cite{Botello-Mendez2011,Habib2012}), functional graphene-based devices (such as transistors~\cite{Schwierz2010} and {\em pn} junctions~\cite{Young2011,Nguyen2013}) and novel state variables (such as pseudospin in mono- and bi-layer graphene, real electronic spin, wavefunction phase~\cite{Habib2012}, and  properties of correlated many-body quantum states like exciton condensates~\cite{Pesin2012}). The greatest roadblock to continued scaling of conventional silicon electronic for digital applications is the power dissipated in various leakage mechanisms~\cite{Keyes2005}. The quantum coherent transport regime and novel state variables in graphene offer a prospect for revolutionary device designs that go beyond traditional (silicon or carbon) field-effect transistors (FETs) based on classical physics. For example, exploiting them would make possible to switch current off without utilizing energy barrier of traditional FETs, thereby evading thermal limitations in the subthreshold regime~\cite{Schwierz2010}.

Concurrently, recent advances in nanofabrication and surface science have pushed the attention towards more complex structures of ``second generation graphene''~\cite{Agnoli2013}, such as chemically modified graphene (i.e., graphene sheets where carbon atoms are replaced by other atoms or entire functional groups) or three-dimensional systems based on the assembly of graphene sheets (where they form interconnected networks or highly complex nanoobjects like hollow nanospheres, crumpled paper, and capsules). Another route  involves creating hybrid nanostructures which combine flat graphene sheets with previously amply explored carbon nanotubes (CNTs) that can be viewed as rolled up sheets of graphene.

For example, the recent experiment~\cite{Jiao2010} has created aligned graphene nanoribbon (GNR) arrays by unzipping of aligned single-walled and few-walled CNT arrays. Then, novel GNR-GNR and GNR-CNT crossbars were fabricated by transferring GNR arrays across GNR and CNT arrays, respectively. The production of such ordered architectures may allow for large scale integration of GNRs into nanoelectronics or optoelectronics. We note that the crossbar geometry has been explored over the past decade using typically semiconductor nanowires, such as those made of Si or GaN~\cite{Lu2008}, as well as crossed CNTs~\cite{Fuhrer2000}. Such motif offers high device density and efficient interconnects between individual devices and functional device arrays, while a reconfigurable crossbar structure allows for effective incorporation of defect-tolerant computing schemes~\cite{Lu2008}.

In contrast to vigorous experimental studies of nanowire crossbars, little theoretical guidance in designing such nanoelectronic devices has been offered by quantum transport simulations. For example, recent studies of GNR-GNR crossbars have utilized semi-empirical tight-binding model~\cite{Botello-Mendez2011} or extended H\"{u}ckel model~\cite{Habib2012} with their parameters fitted to density functional theory (DFT) calculations. These models are then coupled to nonequilibrium Green function (NEGF) formalism~\cite{Stefanucci2013} to compute the zero-bias transmission function~\cite{Botello-Mendez2011}, or even current at finite bias voltage~\cite{Habib2012}. While such approaches can capture charge transfer between different atomic species in equilibrium, they fail to capture more complicated charge redistribution due to nonequilibrium steady transport state driven by the applied bias voltage. On the other hand, the self-consistency of nonequilibrium charge and the corresponding electrostatic (Hartree) potential are indispensable in order to satisfy the gauge-invariant condition according to which {\em I--V} characteristics should not change when electric potential everywhere is shifted by a constant amount~\cite{Christen1996}.

The state-of-the-art approach that has emerged over the past decade and which can capture both charge transfer in equilibrium and charge redistribution in nonequilibrium is NEGF-DFT~\cite{Sanvito2011,Areshkin2010,Taylor2001,Brandbyge2002,Palacios2002,Rocha2006,Stokbro2008}. However, virtually all widely used NEGF-DFT scientific codes, such as commercial ATK~\cite{atk} and NANODCAL~\cite{nanodcal} or open source TranSIESTA~\cite{siesta}, SMEAGOL~\cite{smeagol}, ALACANT~\cite{alacant}, and GPAW~\cite{gpaw} {\em can handle only devices with two electrodes} (GPAW can also simulate multiterminal quantum transport but only in the linear-response regime~\cite{Chen2012b}).

Here we overview the recent extension~\cite{Saha2009,Saha2010a,Saha2010,Saha2011} of NEGF-DFT framework to treat nanojunctions whose active region is attached to more than two electrodes, and then apply it to GNR-CNT crossbars illustrated in Fig.~\ref{fig:fig1}. The primary motivation to investigate four-terminal GNR-CNT crossbars comes from already fabricated structures of this type in Ref.~\cite{Jiao2010}, so that our study could guide selection of specific GNR and CNT elements and their mutual orientation in future efforts to exploit GNR-CNT crossbars for applications in carbon-based nanoelectronics.

Our principal result---{\em I--V} curves plotted in Fig.~\ref{fig:fig4}---offers prediction  of negative differential resistance (NDR)~\cite{Ferry2009} in mGNR-sCNT crossbars. That is, as the source-drain bias voltage increases, current $I_3$ in Fig.~\ref{fig:fig4}(a) reaches a local maximum and then decrease to a valley region before rising again. The conventional NDR-based devices, used in a variety of applications (such as, frequency multipliers, memory, fast switches, and high frequency oscillators up to the THz range), are based on quantum tunneling or intervalley carrier transfer in two-terminal devices like Esaki or Gunn diodes. The advent of graphene has also brought theoretical proposals for NDR in two-terminal GNRs~\cite{Ren2009,Nguyen2013}, as well as experimental demonstration~\cite{Wu2012} of three-terminal large-area graphene where the gate electrode controls the current density and the onset of NDR based on ambipolar transport rather than tunneling. The NDR predicted in Fig.~\ref{fig:fig4}(a), with low onset voltage $V_\mathrm{NDR} \simeq 0.25$ V and peak-to-valley current ratio $\simeq 2.0$, emerges in an ultrascaled geometry containing only $\simeq 460$ carbon atoms in the overlap region.

The paper is organized as follows. In Sec.~\ref{sec:setup}, we discuss details of atomistic structure of three selected GNR-CNT crossbars shown in Fig.~\ref{fig:fig1}. Section~\ref{sec:negfdft} overviews construction of the nonequilibrium density matrix for multiterminal devices as the key quantity that has to be extended in NEGF-DFT framework. In Sec.~\ref{sec:zbtrans}, we discuss zero-bias transmission function between different terminals of GNR-CNT crossbar while current from GNR to CNT at finite bias voltage is discussed in Sec.~\ref{sec:iv}. We conclude in Sec.~\ref{sec:conclusions}.

\section{Device setup for modeling of GNR-CNT crossbars}\label{sec:setup}

We denote graphene nanoribbons with {\em zigzag} edges ZGNRs, while AGNRs denotes nanoribbons with
{\em armchair} edges. We recall that ZGNRs are labeled as $N_z$-ZGNR when they are composed of $N_z$ zigzag chains
of carbon atoms, and $N_a$-AGNR denotes AGNRs composed of $N_a$ dimers of carbon atoms~\cite{Son2006a}.
The single-walled CNTs are labeled by the chiral vector, $\mathbf{C}_h = (n\mathbf{a}_1 + m\mathbf{a}_2)$,
where $n$ and $m$ are integers and $\mathbf{a}_1$ and $\mathbf{a}_2$ are the real space unit vectors of the graphene sheet.
The vector $\mathbf{C}_h$ specifies the way the graphene sheet is wrapped.

In analogy with the experiments on CNT-CNT crossbars~\cite{Fuhrer2000}, which have utilized combination of
metallic and semiconducting CNTs, we select three different setups for GNR-CNT crossbars depicted in panels
(a)--(c) of Fig.~\ref{fig:fig1}: (a) crossed metallic 8-ZGNR and semiconducting single-walled (9,0)-CNT; (b) crossed
semiconducting 14-AGNR and semiconducting single-walled (9,0)-CNT; and (c) crossed metallic 8-ZGNR and single-walled
metallic (5,5)-CNT. The diameter of (9,0)-CNT is $D=0.71$ nm, while $D=0.67$ nm for (5,5)-CNT. The width of 8-ZGNR
is $W=1.78$ nm and the width of 14-AGNR is $W=1.79$ nm.

Simplistic tight-binding models predict that when $n-m$ is a multiple of $3$, subbands will cross at
the Fermi energy implying that CNT is metallic; otherwise, it is expected to be a semiconductor. The energy gap of
semiconducting CNTs is inversely proportional to their diameter. Thus, $(n,n)$-CNTs are expected to always be metallic,
whereas $(n,0)$-CNTs are expected to be metallic only when $n$ is a multiple of $3$. However, experiments and DFT
calculations with properly chosen exchange-correlation (XC) functional show that this is not always correct~\cite{Matsuda2010}.
For example, experiments~\cite{Ouyang2001} have found $E_g = 0.080 \pm 0.005$ eV for (9,0)-CNT, which is well-matched by
the DFT calculations (using the B3LYP flavor of XC functional) $E_g = 0.079$ eV~\cite{Matsuda2010}. The transmission function
plotted in Fig.~\ref{fig:fig3} [see dashed line in subpanels (b) of mGNR-sCNT and sGNR-sCNT panels]  for an infinite homogeneous (9,0)-CNT
exhibits $E_g = 0.09$ eV around the Fermi energy $E_F$.

While simplistic tight-binding models  predict that AGNRs are metallic for $N_a=3n+2$ ($n$ is a positive integer),
DFT calculations show that all AGNRs have energy gap in their electronic subband structure due to transverse confinement~\cite{Son2006a},
which is \mbox{$E_g = 0.14$ eV} in Fig.~\ref{fig:fig3} [see dashed line in subpanel (a) of sGNR-sCNT panel] for the selected 14-AGNR used
for the crossbar in Fig.~\ref{fig:fig1}(b).

Although ZGNRs are insulating at very low temperatures due to one-dimensional
spin-polarized edge states coupled across the width of the nanoribbon, such unusual magnetic ordering and the corresponding
band gap is easily destroyed above $\gtrsim 10$ K~\cite{Yazyev2008,Kunstmann2011}. Thus, they can be considered as good
candidates for metallic electrodes and interconnects~\cite{Areshkin2007a}.
In fact, the recent experiment~\cite{Jia2009}  has confirmed the flow of edge currents~\cite{Chang2012} in metallic ZGNRs which
were actually utilized to increase the heat dissipation around edge defects and, thereby, rearrange atomic structure locally
until sharply defined zigzag edge is achieved.

The infinite GNR and CNT in our simulation are separated initially by a distance \mbox{$d=1.3$ \AA}, where
the overlap region consisting of 486, 480 or 458 carbon and hydrogen atoms [for panels (a)--(c) in Fig.~\ref{fig:fig1}, respectively]
is structurally optimized using VASP simulation package~\cite{Kresse1993,Kresse1996,Kresse1996a}.
Note that central region of the crossbars, for which NEGFs discussed in Sec.~\ref{sec:negfdft} are computed to obtain currents,
includes the GNR-CNT overlap region + portion of the semi-infinite electrodes so that the total number of simulated atoms is 882, 888 or 838
for panels (a)--(c) in Fig.~\ref{fig:fig1}, respectively. The electron-core interactions are described by the projector augmented wave method~\cite{Blochl1994,Kresse1999}, while we use Perdew-Burke-Ernzerhof (PBE)~\cite{Perdew1996} parametrization of the generalized
gradient approximation (GGA) for the XC functional. The cutoff energies for the plane wave basis set used to expand the Kohn-Sham orbitals
are 400 eV for all calculations. A $1 \times 1 \times 1$ $k$-point mesh within Monkhorst-Pack scheme is used for the Brillouin zone integration.
This procedure ensures that Hellmann-Feynman forces acting on ions are less than \mbox{$0.001 \ \mathrm{eV/\AA}$}. The final optimized atomic positions in Fig.~\ref{fig:fig1} show how GNRs curve around CNT within their overlap region, as also observed in the atomic force microscope imaging
of recently fabricated GNR-CNT crossbars~\cite{Jiao2010}. After optimization, the spacing between GNRs and CNTs in Fig.~\ref{fig:fig1}(a)--(c) along the $z$-axis (intersecting with tube axis) is \mbox{$d \approx 3.2$ \AA}.

For the study of near-equilibrium (i.e., linear-response) quantum transport in Sec.~\ref{sec:zbtrans}, we compute zero-bias
transmission coefficients  between different electrodes, labeled as $p \in \{1,2,3,4\}$ in Fig.~\ref{fig:fig1}(d), which are kept at
the same potential. For the study of {\em I--V} characteristics in nonequilibrium transport in
Sec.~\ref{sec:iv}, bias voltage $V_b$ is applied between GNR and CNT by shifting the electrode potentials by $V_1 = V_b/2$, $V_2 = V_b/2$,
$V_3 = - V_b/2$ and $V_4 =-V_b/2$.

\section{NEGF-DFT methodology for multiterminal devices}\label{sec:negfdft}

The traditional CAD tools for electronic device simulations, based on classical drift-diffusion or semiclassical Boltzmann equation, are inapplicable to  quasiballistic nanoscale active region attached to much larger reservoirs. This is exemplified by the crossbars in Fig.~\ref{fig:fig1} which can be
viewed as the nanoscale overlap region of GNR and CNT that is attached to four semi-infinite electrodes which terminate at infinity into macroscopic reservoirs. The proper description of such {\em open quantum systems} can be achieved using quantum master equations for the reduced density matrix of the active region~\cite{Breuer2002,Timm2008} or the NEGF formalism~\cite{Stefanucci2013}. The former is typically used when the active region is weakly coupled to the reservoirs (so that coupling between the active region and the electrodes is treated perturbatively~\cite{Koller2010}), while the latter is employed in the opposite limit.

The NEGF formalism~\cite{Stefanucci2013} for steady-state transport operates with two central quantities, the retarded ${\bf G}(E)$ and the lesser Green functions ${\bf G}^<(E)$, which describe the density of available quantum states and how electrons occupy those states, respectively. Its application to electronic transport often proceeds by combining~\cite{Ferry2009,Datta1995} it with tight-binding (TB) Hamiltonians whose hopping parameters can be fitted using more microscopic theory~\cite{Barraza-Lopez2013}. For example, proper description of quantum transport through GNRs with armchair or zigzag edges requires to
employ TB models with either three orbitals $(p_z,d_{yz},d_{zx})$ per C atom and nearest-neighbor hopping~\cite{Boykin2011}, or single $p_z$ orbital per C atom and third-nearest-neighbor hoppings~\cite{Chang2012}.

However, at finite bias voltage Hamiltonian has to be recomputed self-consistently to capture the nonequilibrium charge redistribution. The NEGF-DFT framework can solve this problem, as long as the coupling between the active device region and the electrodes is strong enough to ensure transparent contact and diminish Coulomb blockade effects~\cite{Koller2010}. The DFT part of  this framework is employed using typical approximations (such as local density approximation, GGA, or B3LYP~\cite{Fiolhais2003}) for its XC functional. The sophisticated computational algorithms~\cite{Sanvito2011,Areshkin2010,Taylor2001,Brandbyge2002,Stokbro2008,Rungger2008} developed to implement the NEGF-DFT framework can be encapsulated by the iterative self-consistent loop
\begin{equation}\label{eq:scloop}
n^{\rm in}({\bf r}) \Rightarrow {\rm DFT} \rightarrow {\bf H}_{\rm KS}[n({\bf r})] \Rightarrow {\rm NEGF} \rightarrow n^{\rm out}({\bf r}).
\end{equation}
The loop starts from the initial input electron density $n^{\rm in}({\bf r})$ and then employs some standard DFT code~\cite{Fiolhais2003}, typically in the basis set of finite-range orbitals for the valence electrons which allows for faster numerics and unambiguous partitioning of the system into the central region and the semi-infinite ideal electrodes. The DFT part of the calculation yields the single particle \mbox{Kohn-Sham} (KS) Hamiltonian
\begin{eqnarray}\label{eq:ks}
\hat{H}_{\rm KS}[n({\bf r})] & = & -\frac{\hbar^2\nabla^2}{2m}  + V^{\rm eff}({\bf r}), \\
V^{\rm eff}({\bf r}) & = & V_H({\bf r}) + V_{\rm xc}({\bf r}) + V_{\rm ext}({\bf r}).
\end{eqnarray}
Here $V^{\rm eff}({\bf r})$ is the DFT mean-field potential due to other electrons, which includes  the Hartree potential $V_H({\bf r})$, the XC potential $V_{\rm xc}({\bf r})$ and the external potential $V_{\rm ext}({\bf r})$.  The inversion of $\hat{H}_{\rm KS}[n({\bf r})]$, represented
in the some basis of local orbitals $\{ \phi_\alpha \}$ as the matrix ${\bf H}_{\rm KS}$ of elements $H_{\alpha \beta} = \langle \phi_\alpha |\hat{H}_{\rm KS}| \phi_\beta \rangle$, yields the retarded Green function ${\bf G}(E)$
\begin{equation}\label{eq:gr}
{\bf G}(E) = \left[ E{\bf S} - {\bf H}_{\rm KS} - \sum_p {\bm \Sigma}_p(E,V_p) \right]^{-1},
\end{equation}
The overlap matrix ${\bf S}$ has elements $S_{\alpha \beta} = \langle \phi_\alpha | \phi_\beta \rangle$. The non-Hermitian matrices ${\bm \Sigma}_p(E,V_p)={\bm \Sigma}_p(E-eV_p)$ are the retarded self-energies due to the ``interaction'' with the semi-infinite electrode $p$ whose electronic structure is assumed to be rigidly shifted by the applied voltage $eV_p$. The self-energies determine escape rates of electrons from the central region into the semi-infinite ideal electrodes, so that an open quantum system can be viewed as being described by the non-Hermitian Hamiltonian ${\bf H}_{\rm open}={\bf H}_{\rm KS}[n({\bf r})] + \sum_p {\bm \Sigma}_p(E,V_p)$. The matrices ${\bm \Gamma}_p(E,V_p)=i[{\bm \Sigma}_p(E,V_p) - {\bm \Sigma}_p^\dagger(E,V_p)]$ account for the level broadening due to the coupling to the electrodes~\cite{Stefanucci2013}.

Assuming the {\em elastic} transport regime (where electron-phonon, electron-electron and electron-spin scattering is neglected), $\mathbf{G}^<(E)$ can be expressed solely in terms of $\mathbf{G}(E)$ to determine the density matrix via
\begin{equation}\label{eq:rho}
{\bm \rho}  =  \frac{1}{2\pi i} \int dE\, \mathbf{G}^<(E) = \frac{1}{2\pi i}  \int dE\, {\bf G}(E) \left[ \sum_p i f_p(E) {\bm \Gamma}_p(E) \right] {\bf G}^\dagger(E).
\end{equation}
The matrix elements $n^{\rm out}(\bf r)=\langle {\bf r} | {\bm \rho} | {\bf r} \rangle$ are the new electron density  as the starting point of the next iteration. This procedure is repeated until the convergence criterion $||{\bm \rho}^{\rm out} - {\bm \rho}^{\rm in}|| < \delta$ is reached, where $\delta \ll 1$ is a suitably chosen tolerance parameter.

The NEGF post-processing of the result of the DFT loop expresses the current flowing into terminal $p$ of the device as
\begin{equation}\label{eq:current}
I_p = \frac{2e}{h} \sum_q \int\limits_{-\infty}^{+\infty} dE\, T_{qp}(E,V_p,V_q) [f_p(E)-f_q(E)].
\end{equation}
Here the transmission coefficients
\begin{equation}\label{eq:trans}
T_{qp}(E,V_p,V_q) = {\rm Tr} \left[ {\bm \Gamma}_q (E,V_q) \cdot {\bf G}(E) \cdot {\bm \Gamma}_{p}(E,V_p) \cdot  {\bf G}^\dagger(E)  \right],
\end{equation}
are integrated over the energy window defined by the difference of the Fermi functions
\begin{equation}\label{eq:fermi}
f_p(E)=\frac{1}{1 + \exp[(E-E_F-eV_p)/k_BT]}
\end{equation}
of macroscopic reservoirs into which semi-infinite ideal electrodes terminate, where $E_F$ is the Fermi energy for the whole device in equilibrium.

The brute force integration in Eq.~\eqref{eq:rho} directly along the real axis can hardly be accomplished due to sharp features (such as van Hove singularities in the density of states) in the integrand which make convergence with increasing number of energy mesh points virtually impossible. Instead, this expression has to be rewritten in the form more suitable for numerical implementation. Thus, the main difference between the usual two-terminal NEGF-DFT methodology~\cite{Brandbyge2002} and the multiterminal~\cite{Saha2009,Saha2010a,Saha2010,Saha2011} one is in computational algorithms employed to construct the density matrix in Eq.~\eqref{eq:rho}, as well as in algorithms for solving the Poisson equation with boundary conditions imposed by the voltages applied to multiple electrodes. This ensures that transmission coefficients $T_{qp}(E,V_p,V_q)$  self-consistently depend on the voltages applied to the electrodes other than $p$, $q$ through the inhomogeneous electrostatic (Hartree) potential acting on electrons within the central device region. The discussion of algorithms for the computation of the density matrix is made transparent by focusing on specific examples where we contrast two-terminal vs. multiterminal cases in Secs.~\ref{sec:twoterminal} and ~\ref{sec:fourterminal}, respectively. We also discuss how to solve the Poisson equation for four-terminal junction in Sec.~\ref{sec:poisson}.

Our MT-NEGF-DFT code~\cite{Saha2009,Saha2010a,Saha2010,Saha2011}, implementing equations discussed in Sec.~\ref{sec:fourterminal}, utilizes  ultrasoft pseudopotentials and PBE~\cite{Perdew1996} parametrization of GGA for the XC functional of DFT. The localized basis set for DFT calculations is constructed from atom-centered orbitals---six per C atom and four per H atom with atomic radius 8.0 Bohr for the devices shown in Fig.~\ref{fig:fig1}. These orbitals are optimized variationally for the electrodes and the central region separately while their electronic structure is obtained concurrently.

\subsection{Two-terminal case}\label{sec:twoterminal}

In the two-terminal case, the integration in Eq.~\eqref{eq:rho} for the elastic transport regime is typically separated into apparent ``equilibrium'' and ``nonequilibrium'' terms as follows~\cite{Brandbyge2002}. We first add and subtract the term $\mathbf{G} {\bm \Gamma}_1 \mathbf{G}^\dagger f_2(E)$ to $\mathbf{ G}^<=\mathbf{G} \left[i f_1(E) {\bm \Gamma}_1 +  i f_2(E) {\bm \Gamma}_2 \right] \mathbf{G}^\dagger$, to get after some rearrangements
\begin{eqnarray}\label{eq:rearrange}
\mathbf{G}^<  =  i \mathbf{G} ({\bm \Gamma}_1 + {\bm \Gamma}_2) \mathbf{G}^\dagger f_2(E)  + i \mathbf{G} {\bm \Gamma}_1 \mathbf{G} [f_1(E) - f_2(E)].
\end{eqnarray}
Here we assume that semi-infinite electrode $2$ is connected to a macroscopic reservoir at the lower electrochemical potential $(E_F + eV_2) < (E_F + eV_1)$.
By substituting the following identity
\begin{eqnarray}\label{eq:identity}
{\bm \Gamma}_1 + {\bm \Gamma}_2 = i \left[ (\mathbf{G}^\dagger)^{-1} - (\mathbf{G})^{-1} \right],
\end{eqnarray}
into Eq.~\eqref{eq:rearrange}, we finally obtain
\begin{equation}\label{eq:im}
\mathbf{G}^<  =  i (\mathbf{G} - \mathbf{G}^\dagger) f_2(E) + i \mathbf{G} {\bm \Gamma}_1 \mathbf{G}^\dagger [f_1(E) - f_2(E)].
\end{equation}
This allows us to rewrite Eq.~\eqref{eq:rho} in the two-terminal case as the sum of two contributions
\begin{equation}\label{eq:rhotwo}
{\bm \rho}   =  -\frac{1}{\pi} \int\limits_\mathrm{EB}^{+\infty} dE \, {\rm Im}\left[ \mathbf{G}(E) \right] f_1(E) +  \frac{1}{2 \pi} \int\limits_{-\infty}^{+\infty}dE \,  \mathbf{G}(E) \cdot {\bm \Gamma}_1(E) \cdot   \mathbf{G}^\dagger(E) \left[ f_1(E) - f_2(E) \right],
\end{equation}
where ${\rm Im}\left[ \mathbf{G}(E) \right] = (\mathbf{G}-\mathbf{G}^\dagger)/2i$. The first ``equilibrium'' term contains integrand which is analytic in the upper complex plane, so that it can be computed via the semicircular path combined with the path in the upper complex plane parallel to the real axis~\cite{Areshkin2010,Brandbyge2002}. The lower energy limit EB is below the bottom valence-band edge. Because $\mathbf{G}(E)$ and $\mathbf{G}^\dagger(E)$ are nonanalytic functions below and above the real axis, respectively, the integrand in the second ``nonequilibrium'' term\footnote{Note that conventionally used ``equilibrium'' and ``nonequilibrium'' terminology for the two terms is handy, but it is not exact since these two terms do not satisfy the gauge-invariant condition, as is the case of proper equilibrium and nonequilibrium density matrices discussed in Ref.~\cite{Mahfouzi2013}.} is nonanalytic function in the entire complex energy plane, so that integration~\cite{Areshkin2010,Sanvito2011} has to be done directly along the real axis between the boundaries around $E_F+eV_1$ and $E_F+eV_2$ determined by the difference of the Fermi functions. The principal challenge for integration in the ``nonequilibrium'' term are sharp peaks in the integrand (due to, e.g., quasibound states or van Hove singularities at the subband edges in the density of state of the electrodes), which proliferate with increasing number of atoms in the central device region and which have to be handled by some version of adaptive mesh of energy points~\cite{Sanvito2011,Areshkin2010}.

\subsection{Multiterminal case}\label{sec:fourterminal}

In the multiterminal case, we use similar manipulation as in Sec.~\ref{sec:twoterminal} to rewrite Eq.~\eqref{eq:rho} as
\begin{eqnarray}\label{eq:rhomulti}
{\bm \rho}^m &=& \frac{1}{2\pi} \int\limits_{-\infty}^{+\infty} dE \, \mathbf{G}(E) {\bm \Gamma}_m(E) \mathbf{G}^\dagger(E) f_m(E)
 + \frac{1}{2\pi} \sum_{p \neq m} \int\limits_{-\infty}^{+\infty} d E \, \mathbf{G}(E) {\bm \Gamma}_p(E) \mathbf{G}^\dagger(E) f_p(E) \nonumber \\
&=& \frac{1}{2\pi} \int\limits_{-\infty}^{+\infty} dE \, \sum_q  \mathbf{G}(E) {\bm \Gamma}_q(E) \mathbf{G}^\dagger(E)  f_m(E)
 + \frac{1}{2\pi} \sum_{p \neq m} \int\limits_{-\infty}^{+\infty} dE \,  \mathbf{G}(E) {\bm \Gamma}_p(E) \mathbf{G}^\dagger(E) f_p(E) \nonumber \\
&& - \frac{1}{2\pi} \sum_{p \neq m} \int\limits_{-\infty}^{+\infty} dE \,  \mathbf{G}(E) {\bm \Gamma}_p(E) \mathbf{G}^\dagger(E) f_m(E) \nonumber \\
&=& - \frac{1}{\pi} \int\limits^{+\infty}_\mathrm{EB} \!\! dE \, \mathrm{Im} \left[\mathbf{G}(E)\right] f_m(E)
 + \frac{1}{2\pi} \sum_{p \neq m} \int\limits_{-\infty}^{+\infty} \!\! dE \, \mathbf{G}(E) {\bm \Gamma}_p(E) \mathbf{G}^\dagger(E) \left[f_p(E) - f_m(E)\right].
\end{eqnarray}
Here energy EB is now chosen below the valence band of all the different electrodes. While in analogy to Sec.~\ref{sec:twoterminal} electrode
$m$ can be chosen as the one which is connected to the reservoir at the lowest electrochemical potential $E_F + eV_m$, in practice we recompute
${\bm \rho}^m$ in Eq.~\eqref{eq:rhomulti} for all possible choices $m=1,2,\ldots,N$ in the given $N$-terminal device. Because of inevitable
errors related to numerical integration, ${\bm \rho}_m$ will not be exactly the same for all $m$. So, in order to minimize such numerical errors,
we construct the final density matrix of a multiterminal device as a weighted average
\begin{equation}\label{eq:rhow}
\rho_{\alpha \beta} = \sum_p w^p_{\alpha \beta} \rho^p_{\alpha \beta}
\end{equation}
where we use notation in terms of the matrix elements $A_{\alpha\beta}=\langle \alpha | \mathbf{A} | \beta \rangle$ in the basis of localized orbitals within the central device region. The weights satisfy $\sum_p w_{\alpha \beta}^p =1$, and should be chosen to minimize the numerical error~\cite{Brandbyge2002,Saha2009}.

For example, for the four-terminal device density matrix elements are given by
\begin{eqnarray}\label{eq:rhofour}
\rho_{\alpha\beta} &=& w_{\alpha\beta}^1\left(\delta_{\alpha\beta}^1 + \Delta_{\alpha\beta}^{12} + \Delta_{\alpha\beta}^{13}
                                                            + \Delta_{\alpha\beta}^{14}\right)
                    + w_{\alpha\beta}^2\left(\delta_{\alpha\beta}^2 + \Delta_{\alpha\beta}^{21} + \Delta_{\alpha\beta}^{23}
                                                            + \Delta_{\alpha\beta}^{24}\right) \nonumber \\
                  &+& w_{\alpha\beta}^3\left(\delta_{\alpha\beta}^3 + \mmat{\Delta}_{\alpha\beta}^{31} + \Delta_{\alpha\beta}^{32}
                                                            + \Delta_{\alpha\beta}^{34}\right)
                  + w_{\alpha\beta}^4\left(\delta_{\alpha\beta}^4 + \Delta_{\alpha\beta}^{41} + \Delta_{\alpha\beta}^{42}
                                                            + \Delta_{\alpha\beta}^{43}\right),
\end{eqnarray}
with weights determined by
\begin{eqnarray}\label{eq:weights}
 w_{\alpha\beta}^1 & = & \frac{(\Delta_{\alpha\beta}^{21})^2 + (\Delta_{\alpha\beta}^{23})^2 + (\Delta_{\alpha\beta}^{24})^2
                                   + (\Delta_{\alpha\beta}^{31})^2 + (\Delta_{\alpha\beta}^{32})^2 + (\Delta_{\alpha\beta}^{34})^2
                                   + (\Delta_{\alpha\beta}^{41})^2 + (\Delta_{\alpha\beta}^{42})^2 + (\Delta_{\alpha\beta}^{43})^2}
                                    {\Delta}, \nonumber \\
 w_{\alpha\beta}^2 & = &  \frac{(\Delta_{\alpha\beta}^{12})^2 + (\Delta_{\alpha\beta}^{13})^2 + (\Delta_{\alpha\beta}^{14})^2
                       + (\Delta_{\alpha\beta}^{31})^2 + (\Delta_{\alpha\beta}^{32})^2 + (\Delta_{\alpha\beta}^{34})^2
                       + (\Delta_{\alpha\beta}^{41})^2 + (\Delta_{\alpha\beta}^{42})^2 + (\Delta_{\alpha\beta}^{43})^2}
                                    {\Delta}, \nonumber \\
 w_{\alpha\beta}^3 & = & \frac{(\Delta_{\alpha\beta}^{12})^2 + (\Delta_{\alpha\beta}^{13})^2 + (\Delta_{\alpha\beta}^{14})^2
                      + (\Delta_{\alpha\beta}^{21})^2 + (\Delta_{\alpha\beta}^{23})^2 + (\Delta_{\alpha\beta}^{24})^2
                      + (\Delta_{\alpha\beta}^{41})^2 + (\Delta_{\alpha\beta}^{42})^2 + (\Delta_{\alpha\beta}^{43})^2}
                                    {\Delta},  \nonumber \\
 w_{\alpha\beta}^4 & = & \frac{(\Delta_{\alpha\beta}^{12})^2 + (\Delta_{\alpha\beta}^{13})^2 + (\Delta_{\alpha\beta}^{14})^2
                      + (\Delta_{\alpha\beta}^{21})^2 + (\Delta_{\alpha\beta}^{23})^2 + (\Delta_{\alpha\beta}^{24})^2
                      + (\Delta_{\alpha\beta}^{31})^2 + (\Delta_{\alpha\beta}^{32})^2 + (\Delta_{\alpha\beta}^{34})^2}
                                    {\Delta},  \nonumber \\
 \Delta  & = &  (N-1) \left[ (\Delta_{\alpha\beta}^{12})^2 + (\Delta_{\alpha\beta}^{13})^2 + (\Delta_{\alpha\beta}^{14})^2
          + (\Delta_{\alpha\beta}^{21})^2 + (\Delta_{\alpha\beta}^{23})^2 + (\Delta_{\alpha\beta}^{24})^2
           + (\Delta_{\alpha\beta}^{31})^2 + (\Delta_{\alpha\beta}^{32})^2 \right. \nonumber \\
           && \left. + (\Delta_{\alpha\beta}^{34})^2 + (\Delta_{\alpha\beta}^{41})^2 + (\Delta_{\alpha\beta}^{42})^2 + (\Delta_{\alpha\beta}^{43})^2 \right].
\end{eqnarray}
Here we use the following shorthand notation
\begin{eqnarray}\label{eq:shorthand}
\delta_{\alpha\beta}^m &  = & - \frac{1}{\pi} \int\limits^{+\infty}_\mathrm{EB} \!\! dE \, \left\{ \mathrm{Im} \left[\mathbf{G}(E)\right]\right\}_{\alpha\beta} f_m(E), \\
\Delta_{\alpha\beta}^{qp} & = & \frac{1}{2\pi} \int\limits_{-\infty}^{+\infty} \!\! dE \, \left\{ \mathbf{G}(E) {\bm \Gamma}_p(E) \mathbf{G}^\dagger(E) \right\}_{\alpha\beta} \left[f_p(E) - f_q(E)\right].
\end{eqnarray}
We test the convergence by increasing the density of the energy mesh, thereby making sure that the integration yields accurate final results.

\subsection{Self-consistent loop for solving the Poisson equation in multiterminal device geometry}\label{sec:poisson}

In a two-terminal system, the initial guess for the electrostatic (Hartree) potential within the central device
region under an applied bias voltage can be chosen simply as a linear interpolation between voltages $V_p$
of the electrodes. However, this step becomes more complicated in multiterminal device geometries. First, one
needs to make sure that the asymptotic potential deep inside of all electrodes will be unaffected by the applied
bias voltage, i.e., the modified potential has to match at the boundary of each electrode and the central
device region. Second, the variation of the potential between any two electrodes through the central region has to be continuous
and uniform. Third, the electrostatic potential in the vacuum region between two arbitrary electrodes has to be realistic.

\begin{figure}
\begin{center}
\includegraphics[scale=0.47,angle=0]{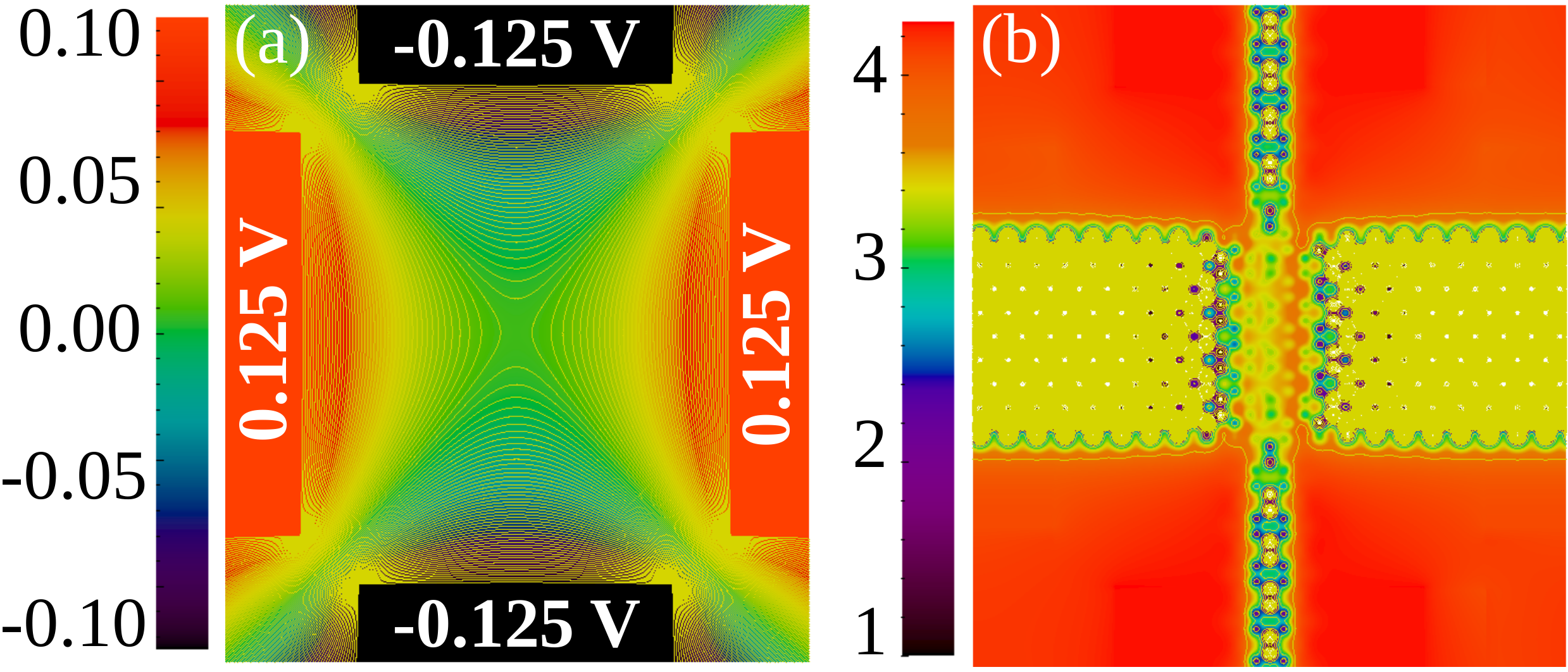}
\end{center}
\caption{(a) Solution of the Laplace equation within the $xy$-planes (of size $5.2 \ \mathrm{nm} \ \times \ 6.0 \ \mathrm{nm}$) of 3D real-space grid of points enclosing mGNR-sCNT crossbar in Fig.~\ref{fig:fig1}(a) whose central region is removed and where voltages applied to the four electrodes provide the boundary conditions. This solution is used as the initial guess for solving the Poisson equation on 3D real-space grid enclosing central region of mGNR-sCNT crossbar, where self-consistent solution within the plane positioned between mGNR and sCNT is shown in panel (b).}
\label{fig:fig2}
\end{figure}

In order to create such an initial potential profile, we iteratively solve the two-dimensional (2D) Laplace equation
\begin{equation}\label{eq:laplace}
\frac{\partial^2 V(x,y,z)}{\partial x^2} + \frac{\partial^2 V(x,y,z)}{\partial y^2}=0
\end{equation}
for a hypothetical system where the central region is empty while using the following boundary conditions: ({\em i}) the initial potential
in the central region is zero or may be the same as the potentials of the four leads, ({\em ii}) the potential in every lead is
unchanged; and ({\em iii}) the potential toward the vacuum region, that is, at the corners of the box, decays. We use the same solution
for each 2D $xy$-plane lined up along the $z$-axis to compose the three-dimensional (3D) real-space grid which encloses the central device region.

This auxiliary solution provides an initial guess for the electrostatic potential profile with boundary conditions set by the voltages
of the leads, which is illustrated in  Fig.~\ref{fig:fig5}(a) for mGNR-sCNT crossbar from Fig.~\ref{fig:fig1}(a) with empty central region.
The Poisson equation for the same mGNR-sCNT crossbar is then updated via the self-consistent loop until the converged solution shown in Fig.~\ref{fig:fig5}(b) is reached. In practice, we find that using solution like the one in Fig.~\ref{fig:fig5}(a) as the first iteration significantly accelerates the convergence. Once the potential and the charge density are converged, the final result is independent of the initial guess.

\section{Zero-bias transmission coefficients of GNR-CNT crossbars}\label{sec:zbtrans}

\begin{figure}
\begin{center}
\includegraphics[scale=0.47,angle=0]{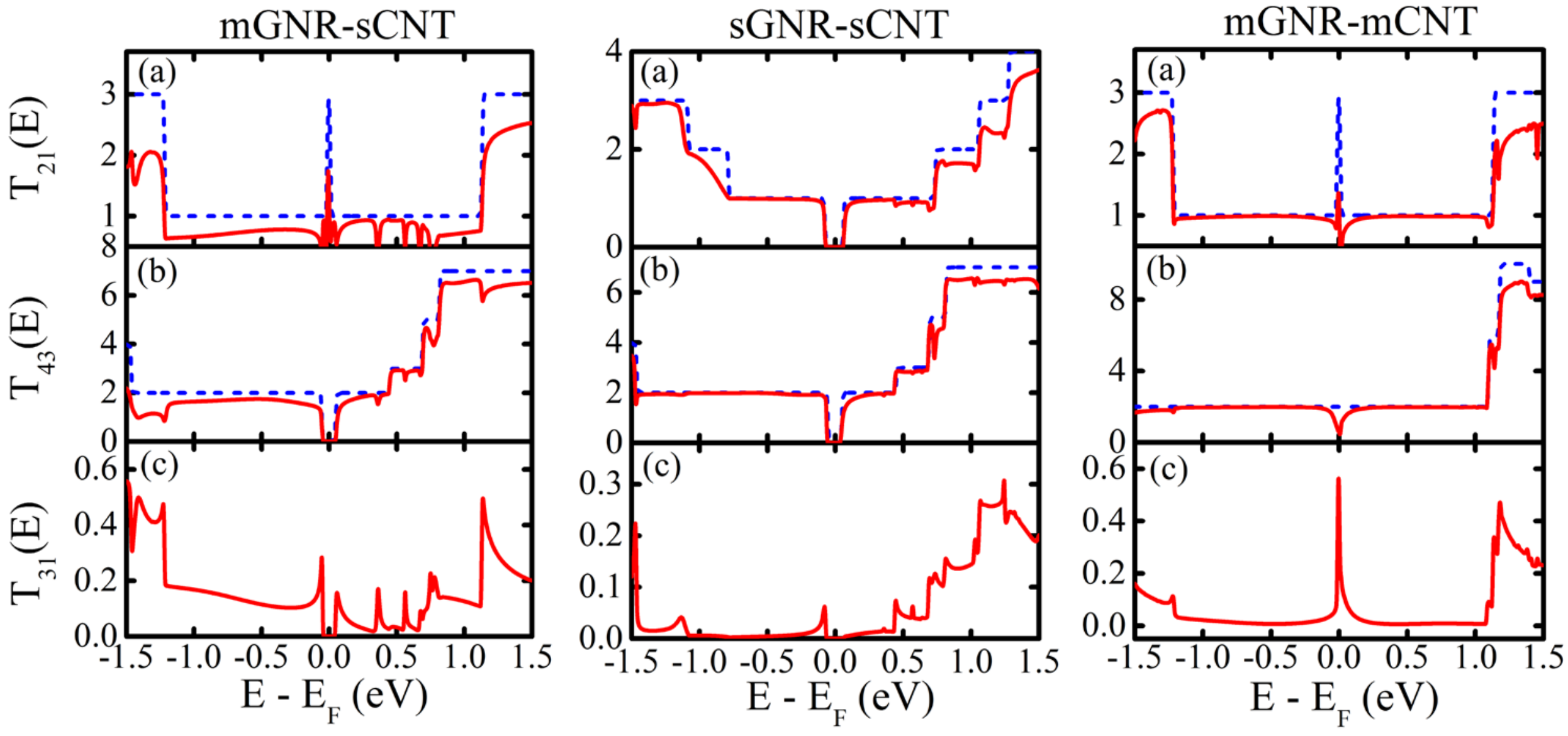}
\end{center}
\caption{Solid lines plot selected zero-bias transmission coefficients $T_{pq}(E)$ between electrodes [labeled in Fig.~\ref{fig:fig1}(d)] of mGNR-sCNT, sGNT-sCNT and mGNR-mCNT crossbars depicted in Fig.~\ref{fig:fig1}. Dashed lines in subpanels (a) and (b) plot quantized zero-bias transmission of individual GNRs or CNTs, respectively, that form the crossbar denoted on the top of each three-panel figure.}
\label{fig:fig3}
\end{figure}

In this section, we consider quantum transport quantities that are relevant for the description of
the linear-response regime in crossbars depicted in Fig.~\ref{fig:fig1}. In that case, DFT Hamiltonian computed in equilibrium yields the transmission coefficients in Eq.~\eqref{eq:trans}
where all $V_p=0$. The integration of the transmission coefficients $T_{qp}(E)$ yields the conductance coefficients at finite
temperature
\begin{equation}\label{eq:cond}
G_{qp} = \frac{2e^2}{h} \int dE\, T_{qp}(E) \left(-\frac{ \partial f}{\partial E} \right),
\end{equation}
where $T_{qp}(E)=T_{pq}(E)$ and $G_{qp} = G_{pq}$ in the absence of external magnetic field~\cite{Datta1995,Buttiker1986}. The conductance coefficients connect current in electrode $p$ to voltages applied to electrodes $q \neq p$ via the multiterminal Landauer-B\"{u}ttiker formula~\cite{Ferry2009,Datta1995,Buttiker1986}
\begin{equation}\label{eq:lb}
I_p = \sum_q G_{qp} [V_p-V_q].
\end{equation}
which is the linear-response limit simplification of Eq.~\eqref{eq:current}.

The zero-bias transmission coefficients along GNR, CNT and between GNR and CNT for three devices in Figs.~\ref{fig:fig1} are shown in
Fig.~\ref{fig:fig3}. In addition, dashed line in subpanels (a) in each of these three Figs. shows quantized zero-bias transmission function of an
isolated infinite homogeneous GNR attached to two macroscopic reservoirs, while dashed line in subpanels (b) shows the same information for an
isolated infinite homogeneous CNT. This allows one to visually estimate the impact of the proximity of GNR on the linear-response transport properties of
CNT and vice versa. In all cases plotted in Fig.~\ref{fig:fig3} we find sizable transmission coefficient $T_{31}(E)$ which signifies non-tunneling
nature of transport between crossed GNR and CNT in junctions shown in Fig.~\ref{fig:fig1}.

\section{Current-voltage characteristics of GNR-CNT crossbars}\label{sec:iv}

\begin{figure}
\begin{center}
\includegraphics[scale=0.3,angle=0]{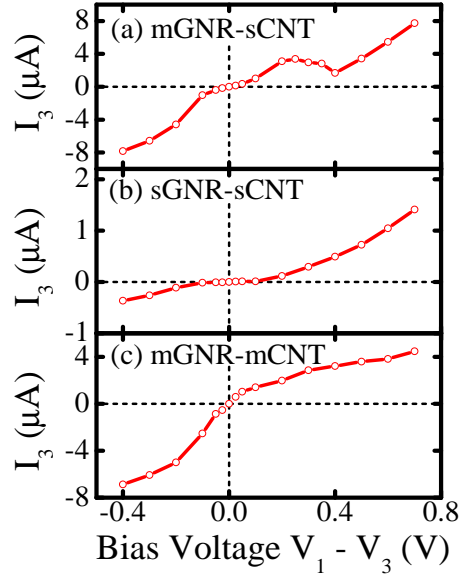}
\end{center}
\caption{Current in electrode 3 [for electrode labels see Fig.~\ref{fig:fig1}(d)] versus bias voltage $V_b=V_1 -V_3$ applied between electrodes 1 and 3 (the same bias voltage $V_b=V_2 - V_3$ is also applied between electrodes 2 and 3) for three GNR-CNT crossbars depicted in Fig.~\ref{fig:fig1}. The NDR in panel (a) is characterized by the onset voltage $V_\mathrm{NDR} \simeq 0.25$ V and peak-to-valley current ratio $\mathrm{PVCR} \simeq 2.0$.}
\label{fig:fig4}
\end{figure}

In this section we consider {\em nonequilibrium} quantum transport in CNT electrode 3 of devices depicted in Fig.~\ref{fig:fig1}, which is driven by
finite bias voltage and, therefore, far outside the linear-response regime discussed in Sec.~\ref{sec:zbtrans}. The bias voltage $V_b=V_1-V_3=V_2-V_3$
is applied between electrodes 1 and 3, as well as 2 and 3, labeled in Fig.~\ref{fig:fig1}(d). This means that current $I_3$ plotted in Fig.~\ref{fig:fig4}
has contribution from two electron fluxes, $1 \rightarrow 3$ and $2 \rightarrow 3$, which are summed via Eq.~\eqref{eq:current}. We use convention that
current is positive in electrode $p$ if it flows into this electrode (i.e., electrons are flowing out of electrode 3 in Fig.~\ref{fig:fig4} for positive bias voltage $V_b >0$). We use $T=300$ K in Eq.~\eqref{eq:fermi} while neglecting any inelastic scattering processes that would add additional self-energies into Eq.~\eqref{eq:gr}, thereby making usage of NEGF-DFT beyond $\lesssim 100$ atomic orbitals prohibitively computationally expensive~\cite{Frederiksen2007}.

Unlike experiments on CNT-CNT crossbars~\cite{Fuhrer2000}, where intertube current (i.e., equivalent of our current $I_3$) was measured $\simeq 0.2$ $\mu$A at
forward bias voltage $\simeq 0.8$ V in mCNT-sCNT device (at $T=100$ K), the current $I_3$ in all three of our crossbars is an order of magnitude larger at similar bias voltage. This is due to the overlap of orbitals on C atoms of GNR and CNT (see Fig.~\ref{fig:fig6}) which introduces non-negligible effective hopping between them, so that transport is not governed solely by quantum tunneling through the vacuum gap as in CNT-CNT crossbars.

\begin{figure}
\begin{center}
\includegraphics[scale=0.3,angle=0]{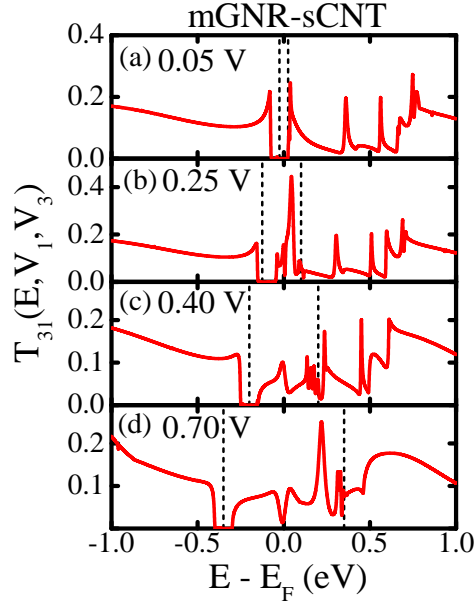}
\end{center}
\caption{Transmission coefficient $T_{31}(E,V_1,V_3)$ at finite bias voltage $V_1 -V_3$. Panel (b) corresponds to the peak current and panel (c) to the valley current of NDR displayed in Fig.~\ref{fig:fig4}, while panels (a) and (d) are outside of the NDR region. Vertical dashed lines in each panels enclose the bias voltage window, within which $T_{31}(E,V_1,V_3)$ is integrated to get one of the two contributions [the other one is governed by the coefficient $T_{32}(E,V_2,V_3)$] to current $I_3$.}
\label{fig:fig5}
\end{figure}

The most interesting feature of $I_3$ vs. $V_b$ curves in Fig.~\ref{fig:fig4} is NDR shown in Fig.~\ref{fig:fig4}(a) for mGNR-sCNT crossbar. The applications of devices exhibiting NDR in microwave, switching, and memory devices requires low onset voltage $V_\mathrm{NDR}$ and a high peak-to-valley current ratio (PVCR) to reduce power consumption and improve the efficiency. For example, a key requirement of NDR devices for use in embedded memory applications is a low valley current to reduce the standby power consumption while concurrently having a sufficient peak current to charge the parasitic node capacitance. The NDR exhibited by mGNR-sCNT crossbar in Fig.~\ref{fig:fig4}(a) has parameters $V_\mathrm{NDR} \simeq 0.25$ V and  $\mathrm{PVCR} \simeq 2.0$.

We can compare this with three-terminal (based on FET configuration) large-area graphene NDR device recently fabricated by the IBM team~\cite{Wu2012}
which exhibited $V_\mathrm{NDR} \simeq 1.3$ V and $\mathrm{PVCR} \simeq 1.1$. The onset of NDR behavior $V_\mathrm{NDR}$ can be shifted by the gate
voltage in this device, but it remains above 1 V. NDR was also observed~\cite{Bushmaker2009} in suspended quasi-metallic single-walled CNTs where
applying a gate voltage switches {\em I--V} characteristics from Ohmic behavior to nonlinear behavior with $V_\mathrm{NDR} \simeq 0.9$ V and
$\mathrm{PVCR} \simeq 1.05$. Previously explored semiconductor devices with low $V_\mathrm{NDR} = 0.1$ V and high $\mathrm{PVCR}=6.2$ include trench-type InGaAs/InAlAs quantum-wire-FET at 40 K in a simple three-terminal configuration in which NDR can also be controlled by the gate voltage~\cite{Jang2003}.

The origin of NDR in this system of noninteracting quasiparticles is explained in Fig.~\ref{fig:fig5} which plots the transmission coefficient $T_{31}(E,V_1,V_3)$ as a function of the bias voltage. At the onset of NDR, transmission peak enters the bias voltage window in Fig.~\ref{fig:fig5}(b), while at the NDR valley transmission coefficient within the larger bias window is reduced in Fig.~\ref{fig:fig5}(c). Additional information is provided by the local density of states (LDOS) at finite bias voltage which is plotted in Fig.~\ref{fig:fig6}. The LDOS becomes enhanced at NDR peak both along the edges of ZGNR and in its interior.

\begin{figure}
\begin{center}
\includegraphics[scale=0.47,angle=0]{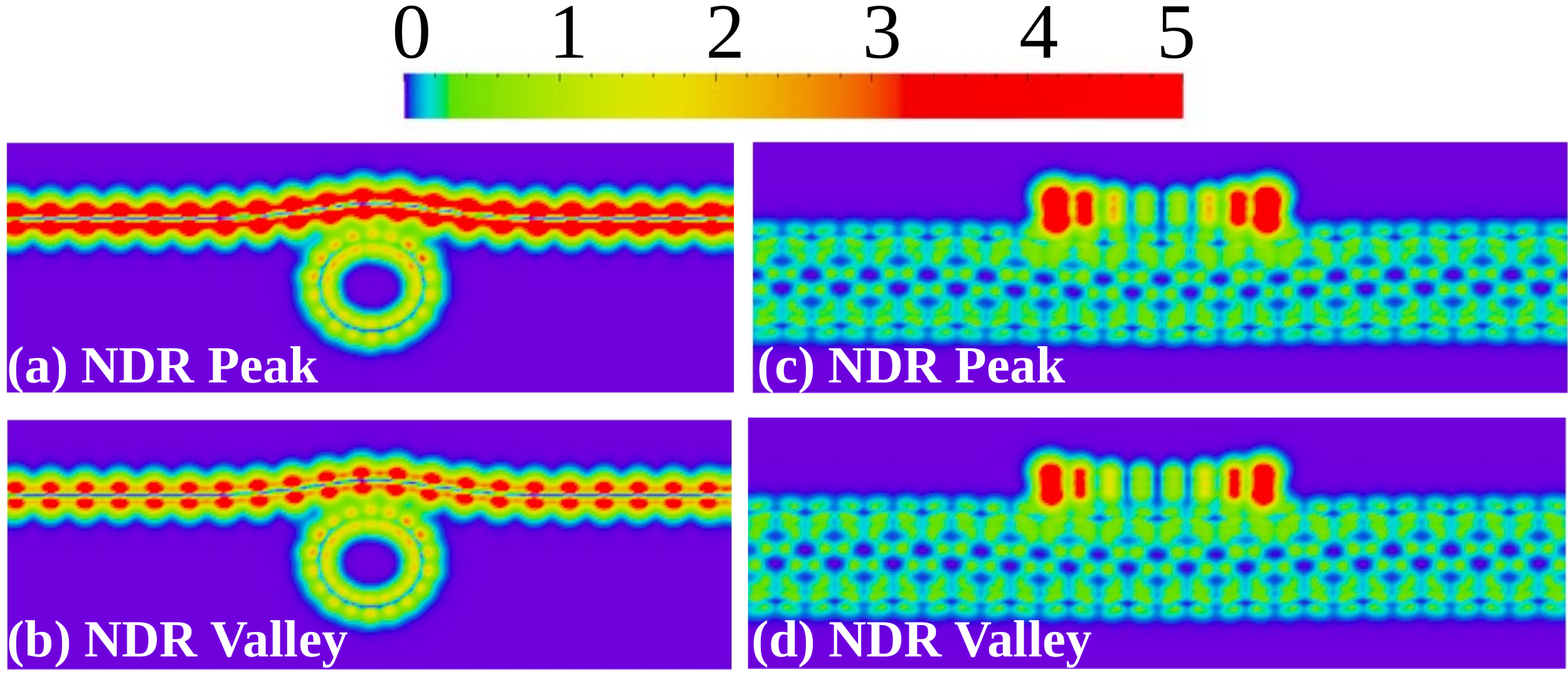}
\end{center}
\caption{The local density of states (LDOS) for mGNR-sCNT crossbar depicted in Fig.~\ref{fig:fig1}(a) at finite bias voltage: (a) and (c) $V_b=V_1-V_3=0.25$ V, which corresponds to the peak of NDR in Fig.~\ref{fig:fig4}(a); and (b) and (d) $V_b = V_1-V_3 = 0.4$ V, which corresponds to the valley of NDR in Fig.~\ref{fig:fig4}(a). LDOS is integrated within the interval [-0.05, \ 0.1] eV in panels (a) and (c) or [-0.15, \ 0.2] eV in panels (b) and (d). These two-dimensional plots are obtained by summing LDOS over all $xz$-planes (along the $y$-axis) which pass through the overlap region of mGNR and sCNT, or over all $yz$-planes (along the $x$-axis) passing through the overlap region. The physical size of images plotted in panels (a) or (b) is  $5.2 \ \mathrm{nm} \ \times \ 1.9 \ \mathrm{nm}$, while in panels (c) or (d) it is $6.0 \ \mathrm{nm} \ \times \ 1.9 \ \mathrm{nm}$.}
\label{fig:fig6}
\end{figure}

\section{Concluding remarks} \label{sec:conclusions}
In conclusion, using the NEGF-DFT framework for first-principles quantum transport modeling, which was originally developed for two-terminal devices and recently extended to multiterminal ones, we have analyzed current between GNR and single-walled CNT where electrons traverse the vacuum gap between them. The GNR covers CNT over a nanoscale region where their overlap contains only $\simeq 460$ carbon and hydrogen atoms. Following up on the recent experimental fabrication of GNR-CNT crossbars~\cite{Jiao2010}, as well as previous experiments on CNT-CNT crossbars~\cite{Fuhrer2000}, we have investigated three examples of GNR-CNT crossbars combining metallic and semiconducting crossed nanowires.

While all three types of four-terminal GNR-CNT crossbars exhibit nonlinear current-voltage {\em I--V} characteristics, which is asymmetric with respect to changing the bias voltage from positive to negative, mGNR-sCNT crossbar also exhibits NDR with low onset voltage $V_\mathrm{NDR} \simeq 0.25$ V and
peak-to-valley current ratio $\simeq 2.0$. The origin of NDR is examined by looking at the transmission coefficient between GNR and CNT semi-infinite electrodes and local density of states at finite bias voltage.

We also provide an overview of essential computational issues that have to be tackled when extending NEGF-DFT framework to nanojunctions containing more that two electrodes~\cite{Saha2009,Saha2010a,Saha2010,Saha2011}.

%

\begin{acknowledgements}
We thank  V. Meunier  for illuminating discussions. Financial support from NSF under Grant No. ECCS 1202069 is gratefully acknowledged. The supercomputing time was provided in part by NSF through XSEDE resource TACC Stampede.
\end{acknowledgements}



\end{document}